\documentclass[prb,aps,twocolumn,showpacs,floats,amssymb,amsfonts]{revtex4}

\usepackage{graphicx}
\usepackage{bm}

\newcommand{\virtX}{\langle X \rangle}
\newcommand{\virtY}{\langle Y \rangle}

\newcommand{\bDV}{\overline{\Delta V}}
\newcommand{\bDU}{\overline{\Delta U}}

\newcommand{\bDUs}{\overline{\Delta U}}

\begin{document}

\title{
Schottky barrier heights at polar metal/semiconductor interfaces
}

\author{C. Berthod}
\affiliation{DPMC, Universit\'e de Gen\`eve, 24 quai Ernest-Ansermet,
CH--1211 Gen\`eve 4, Switzerland}

\author{N. Binggeli}
\affiliation{International Center for Theoretical Physics (ICTP) and
INFM DEMOCRITOS National Center, Strada Costiera 11, I--34014 Trieste, Italy}

\author{A. Baldereschi}
\affiliation{
Institut Romand de Recherche Num\'erique en Physique des Mat\'eriaux (IRRMA),
CH--1015 Lausanne, Switzerland
}

\date{\today}

\begin{abstract}

Using a first-principle pseudopotential approach, we have investigated the
Schottky barrier heights of abrupt Al/Ge, Al/GaAs, Al/AlAs, and Al/ZnSe (100)
junctions, and their dependence on the semiconductor chemical composition and
surface termination. A model based on linear-response theory is developed,
which provides a simple, yet accurate description of the barrier-height
variations with the chemical composition of the semiconductor. The larger
barrier values found for the anion- than for the cation-terminated surfaces are
explained in terms of the screened charge of the polar semiconductor surface
and its image charge at the metal surface. Atomic scale computations show how
the classical image charge concept, valid for charges placed at large distances
from the metal, extends to distances shorter than the decay length of the
metal-induced-gap states.

\end{abstract}

\maketitle

\section{Introduction}

Metal/semiconductor (MS) interfaces have been the focus of extensive
theoretical and experimental studies for several
decades.\cite{Rhoderick-88,Brillson-92,Tung-92,Margaritondo-99} To date,
however, we are still far from a complete understanding of the factors which
control the Schottky barrier height (SBH) at these interfaces. In recent years,
new research activities have been developed in the area of band engineering at
MS
interfaces\cite{Franceschi-98,Sorba-96,Orto-94,Cantile-94,Koyanagi-93,Costa-91}
and on the properties of metal/wide-gap-semiconductor
contacts.\cite{Lazzarino-98,Chen-94,Barinov-01} These developments have
stimulated renewed interest in some basic issues concerning Schottky barriers,
and in particular in the mechanisms that control the SBH dependence on
bulk-semiconductor and interface-specific characteristics.

The problem of Schottky barrier formation has been traditionally addressed by
studying the dependence of the SBH on the metal used in the
junction.\cite{Monch-90} Early studies suggested a Schottky-Mott behavior
controlled by the metal work function for highly ionic or wide-gap
semiconductors, and a weak dependence on the metal type and on the junction
fabrication method for the most covalent semiconductors, such as Si or
GaAs.\cite{Brillson-92,Monch-90} The latter trend was generally attributed to
various Fermi level pinning mechanisms, such as pinning by metal-induced-gap
states (MIGS)\cite{Heine-65} at an intrinsic charge neutrality level of the
semiconductor\cite{Tersoff-84,Flores-87} or pinning by native defect states of
the semiconductor at some extrinsic gap level.\cite{Spicer-93,Woodall-82}
Furthermore, a correlation between Schottky barriers and heterojunction band
offsets was observed experimentally for a number of
systems,\cite{Margaritondo-87} and similarly ascribed to Fermi level pinning at a
bulk reference level. Finally, the effect of the semiconductor
ionicity on the SBH trend with the metal work function was examined in
pioneering self-consistent studies of jellium/semiconductor contacts,
and the trend could also be generally understood in terms of MIGS
properties of the semiconductor.\cite{Louie-77}

More recent experiments on metal contacts to covalent semiconductors, however,
have revealed a much weaker electronic pinning than was previously
believed.\cite{Brillson-92} In particular, there have been reports of
considerable changes in metal/Si and metal/GaAs SBH's obtained by altering the
structural properties and/or the chemical composition of the
interface.\cite{Sullivan-93,Cantile-94} The conclusion that the SBH does depend
most generally on the microscopic atomic structure of the interface has been
reached by many authors, both on experimental\cite{Brillson-92,Tung-92,%
Franceschi-98,Sorba-96,Orto-94,Cantile-94,Lazzarino-98,Chen-94,Sullivan-93} and
theoretical\cite{Lazzarino-98,Zhang-85,Dandrea-93,Charlesworth-94,Berthod-96,%
Ruini-97,Bardi-99} grounds. While opening a promising line of research on
Schottky barrier engineering, these observations complicate seriously the
search for simple models of Schottky barrier formation, since the inclusion of
the interface atomic structure seems unavoidable.

Given the complexity and variety of the atomic structure at metal/semiconductor
contacts, it seems unlikely that a simple unified model could emerge and
entirely cover the various facets of Schottky barrier formation. Conversely, a
systematic investigation of the problem starting from abrupt, defect free
interfaces, and progressively introducing perturbations at the interface could
help identifying relevant microscopic mechanisms and provide a firmer basis for
modeling Schottky barrier properties. Progress in computational physics have
made possible accurate {\it ab initio\/} calculations of the electronic
structure of MS contacts, and the complexity of the systems which can be
examined is steadily increasing; this type of computations can provide the means
to carry out such an investigation and probe the correlation between microscopic
atomic structures and SBH's. The present study is a first step in this
direction.

In this article, we study from first principles the dependence of the SBH on
selected bulk and surface characteristics of the semiconductor, for a given
metal. Specifically, we examine abrupt Al/$X$~(100) junctions, where
$X=$ (Ge, GaAs, AlAs, and ZnSe) are lattice matched semiconductors
of increasing ionicity, and we investigate the microscopic mechanisms
responsible for the SBH changes with the semiconductor chemical
composition and surface termination (cation or anion).
A model based on a linear-response-theory scheme is then developed,
which explains our {\it ab initio\/} results and SBH trends observed
experimentally.

\section{Method of calculation}

We have carried out {\it ab initio\/} calculations, within the Local-Density
Approximation (LDA) to Density Functio\-nal Theory (DFT), using the
pseudopotential method.\cite{Pickett-89} We used norm-conserving
scalar-relativistic Troullier-Martins
pseudopotentials\cite{Troullier-91,Pseudo} in the Kleinman-Bylander non-local
form\cite{Kleinman-82} and the exchange-correlation functional of Ceperley and
Alder.\cite{Ceperley-80} The electronic states were expanded on a plane-wave
basis set using a kinetic energy cut-off of 20~Ry. We used supercells
containing 7 Al layers and 13 semiconductor layers ($7+13$ supercell) to model
defect-free Al/$X$~(100) junctions. In section~\ref{sec:termination}, we
employed larger supercells ($7+21$) to investigate the screening of
substitutional charges placed in the junctions, and to compute the parameters
$(D_s,\,\delta_s)$ necessary to model this screening. All supercell
calculations were performed with a (2,\,6,\,6) Monkhorst-Pack $k$-point
grid.\cite{Monkhorst-76}

\begin{figure}[b!]
\includegraphics[width=8.6cm]{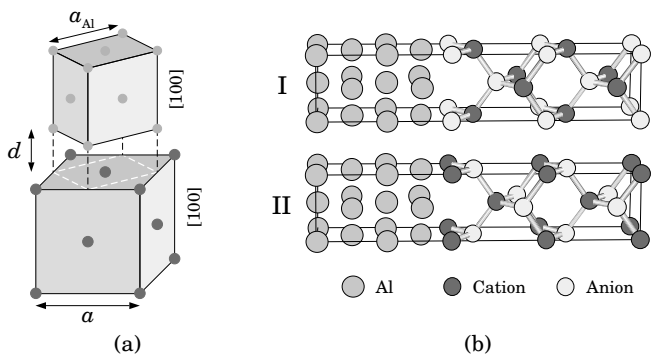}
\caption{\label{fig:structure}
(a) Epitaxial alignment of Al on the (100) surface of zinc-blende
semiconductors verifying the lattice-matching condition
$a_{\text{Al}}=a/\sqrt{2}$. (b) Atomic structure of the abrupt
Al/semiconductor~(100) interface. The semiconductor surface is terminated
either by an anion (I) or a cation (II) plane.}
\end{figure}

We considered ideally abrupt epitaxial junctions and neglected atomic
relaxation at the interfaces. The effect of atomic relaxation at the
Al/GaAs~(100) and Al/ZnSe~(100) interfaces has been examined in
Refs.~\onlinecite{Lazzarino-98} and \onlinecite{Berthod-98}. Atomic relaxation
decreases (increases) the $p$-type Schottky barriers of the abrupt Al/GaAs
(Al/ZnSe) junctions by 0--0.1~eV\cite{Berthod-98}
(0.1--0.2~eV\cite{Lazzarino-98,Berthod-99}), and has no influence on the SBH
ordering of the anion- and cation-terminated interfaces.

The epitaxial alignment of Al on the (100) surfaces of the four semiconductors
under study is illustrated in Fig.~\ref{fig:structure}(a). This type of
alignment corresponds to the lattice-matching condition:
$a_{\text{Al}}=a/\sqrt{2}$, where $a$ is the semiconductor lattice parameter.
The Al [100] direction is parallel to the semiconductor [100] axis, and the
whole Al fcc lattice is rotated by 45$^\circ$ about its [100] axis with respect
to the semiconductor substrate. Experimentally --- and also in our calculations
--- Ge, GaAs, AlAs, and ZnSe are lattice-matched semiconductors, and the
equilibrium lattice constant of Al is slightly larger (1\%) than $a/\sqrt{2}$.
This results in a small compressive strain in the Al in-plane lattice parameter,
which is accommodated by an elongation ($\sim 3$\%) of the Al overlayer,
assuming pseudomorphic conditions. For the semiconductor lattice parameter, we
used the theoretical value $a=5.55$~\AA\ ($a^{\text{exp.}} = 5.65$~\AA). The
metal-semiconductor interlayer distance $d$ at the junction was taken as the
average between the (100) interlayer spacings in the semiconductor and in the
(strained) Al bulk parts, i.e., $d=1.72$~\AA. The polar Al/$X$~(100) junction
offers two inequivalent abrupt interfaces, either with anion- or
cation-terminated semiconductor surface, which we both considered in our study
[see Fig.~\ref{fig:structure}(b)]. In what follows, we will refer to the anion-
and to the cation-terminated interface as interface~I and II, respectively.

To evaluate the $p$-type SBH, $\phi_p$, we used the same approach as in
previous studies:\cite{Berthod-96,Berthod-98}
	\begin{equation}\label{eq:phi_p}
		\phi_p = \Delta V + \Delta E_p,
	\end{equation}
where $\Delta V$ is the electrostatic-potential lineup at the interface and
$\Delta E_p$ is the difference between the Fermi level in the metal and the
valence-band maximum (VBM) in the semiconductor, each measured with respect to
the average electrostatic potential in the corresponding crystal. The
band-structure term $\Delta E_p$ is characteristic of the individual bulk
crystals forming the junction. This term was computed using the Kohn-Sham (KS)
eigenvalues of standard bulk band-structure calculations. The potential lineup
$\Delta V$ contains all interface-specific contributions to $\phi_p$ and was
obtained --- via Poisson's equation and using a macroscopic average
technique\cite{Berthod-96,Berthod-98} --- from the self-consistent supercell
charge density.

For a meaningful comparison of our calculated SBH's with experiment, $\Delta
E_p$ should include quasiparticle and spin-orbit corrections. The spin-orbit
correction is simply $+\frac{\Delta_{\text{so}}}{3}$, where
$\Delta_{\text{so}}$ is the total spin-orbit splitting at the semiconductor
valence-band maximum, which was taken from experiment. For a metal, in
principle, the {\it exact\/} KS Fermi energy and the quasiparticle Fermi energy
must coincide at zero temperature.\cite{Dreizler-90} Furthermore, LDA
calculations for the work functions of various Al surfaces performed with the
same method and the same pseudopotentials as in the present study --- and
neglecting many-body corrections on the Al Fermi energy --- yielded values
which agree with the experimental data to within a few tenths of
meV.\cite{Fall-98} In the present study, therefore, for the metal Fermi energy
we just used the LDA result. The corrected band term is thus $\Delta E_p=\Delta
E_p^{\text{KS}}-\Delta E_{\text{qp}}-\frac{\Delta_{\text{so}}}{3}$, where
$\Delta E_p^{\text{KS}}$ is the KS band term and $\Delta E_{\text{qp}}$ is the
difference between the quasiparticle and KS semiconductor VBM energies.

For the quasiparticle corrections, we used the results of GW calculations
taken from the literature. For Al/GaAs we used the correction $\Delta
E_{\text{qp}}^{\text{GaAs}}=-0.36$~eV evaluated by Charlesworth {\it et
al.},\cite{Charlesworth-94} who employed for the reference LDA calculations the
same exchange-correlation potential as we do. For Ge (AlAs) we used the
correction for GaAs, and the difference between the Ge (AlAs) and GaAs
corrections evaluated in Ref.~\onlinecite{Zhu-91}, i.e., $\Delta
E_{\text{qp}}^{\text{Ge}}-\Delta E_{\text{qp}}^{\text{GaAs}}=+0.09$~eV ($\Delta
E_{\text{qp}}^{\text{AlAs}}-\Delta
E_{\text{qp}}^{\text{GaAs}}=-0.11$~eV).\cite{GWxc} The quasiparticle corrections
to the band structure of ZnSe have been evaluated in
Ref.~\onlinecite{Zakharov-94}. As the LDA bandgap in our calculations and in
Ref.~\onlinecite{Zakharov-94} are different, due to the different
pseudopotentials employed, we took the valence-band-edge correction of
Ref.~\onlinecite{Zakharov-94} and scaled it by the ratio of the difference
between the LDA and GW bandgap in the two calculations. The resulting estimate
for $\Delta E_{\text{qp}}^{\text{ZnSe}}$ is $-0.50$~eV. Using the experimental
spin-orbit splittings $\Delta_{\text{so}}^X=$~0.30, 0.34, 0.28, and 0.43~eV for
$X=$ Ge, GaAs, AlAs, and ZnSe,\cite{Landolt-82} the total corrections are 0.17,
0.25, 0.36, and 0.36~eV, respectively.\cite{erratum} The numerical uncertainty on
the absolute value of the SBH's is estimated as $\sim 0.1$~eV for Al/Ge,
Al/GaAs, and Al/AlAs, and as $\sim 0.2$~eV for Al/ZnSe. For a given interface
geometry, however, the relative barrier values ($\Delta\phi_p$ in
Table~\ref{tab:delta_phi_p}) are considerably more accurate, i.e, have an
estimated numerical accuracy of $\sim50$~meV.

\section{Results for the Schottky barrier heights}

\begin{figure}[t!]
\includegraphics[width=8cm]{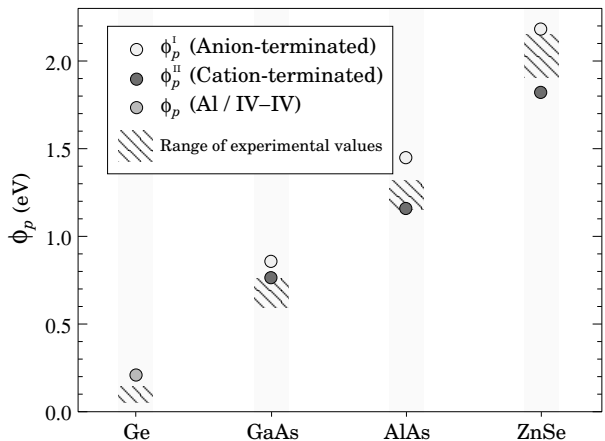}
\caption{\label{fig:phi_p-exp}
Schottky barrier heights (SBH's) at Al/$X$~(100) contacts, $X=$ Ge, GaAs, AlAs,
and ZnSe. The circles show the calculated SBH's for the ideal anion- and
cation-terminated junctions, corrected for quasiparticle and spin-orbit
effects. The shaded regions show the ranges of experimental values (see text).}
\end{figure}

The calculated SBH's for the abrupt Al/$X$~(100) interfaces, including
many-body and spin-orbit corrections, are given in Table~\ref{tab:phi_p}. We
observe a systematic difference between the type-I and type-II interfaces: the
$p$-type SBH is always higher for the type-I (anion-terminated) interface. This
difference increases with increasing semiconductor ionicity. Our theoretical
results are compared with experimental SBH values in Fig.~\ref{fig:phi_p-exp}.
For the Al/Ge, Al/GaAs, and Al/AlAs systems the experimental ranges correspond
to data obtained by transport measurements. In the case of Al/GaAs,
photoemission measurements --- performed at low metal coverage --- give rise to
a wider range of SBH values,\cite{Brillson-93} but the scattering in the data
decreases significantly when thick metallic overlayers are deposited and the
barriers are measured by transport techniques. For Al/ZnSe we are not aware of
any transport data and we used photoemission results.

In the case of Al/Ge, no SBH measurement has been performed, to our knowledge,
on the (100)-oriented interface. In Fig.~\ref{fig:phi_p-exp} we have thus used
the existing transport data\cite{Thanailakis-73} on Al/$n$-Ge~(111) junctions
($\phi_n=0.52$--0.61~eV), together with the Ge experimental bandgap at room
temperature, $E_g^{\text{Ge}} = 0.66$~eV.\cite{Sze-81} The resulting barrier
heights $\phi_p = 0.05$--0.14~eV compare reasonably well with our calculated
value of 0.21~eV. In the case of Al/GaAs~(100) the transport measurements give
values of $\phi_p$ between 0.58~eV and
0.76~eV\cite{Cantile-94,Koyanagi-93,Costa-91,Johnson-76,Cho-78,Wang-83,%
Barret-83,Eglash-84,Waldrop-85,Missous-86,Eglash-87,Missous-90,Bosacchi-94} (we
used $E_g^{\text{GaAs}}=1.42$~eV to estimate the $p$-type barrier heights from
measurements performed on Al/$n$-GaAs junctions). This is in relatively good
agreement with our calculated SBH of 0.76~eV for the Ga-terminated interface,
and still consistent with our value of 0.86~eV for the As-terminated interface.
Concerning the effect of the GaAs-surface stoichiometry on the measured SBH, we
note that different conclusions have been reached by different groups. Some
studies, including Refs.~\onlinecite{Cho-78} and \onlinecite{Wang-83}, have
found a small ($\sim 0.1$~eV) difference between the SBH's measured in junctions
fabricated on As-rich and on Ga-rich surfaces (As-rich leading to higher
$\phi_p$, consistent with our results), while other studies, such as
Refs.~\onlinecite{Barret-83} and \onlinecite{Missous-86}, found no difference.

\begin{table}[t!]
\caption{\label{tab:phi_p}
Estimated quasiparticle and spin-orbit corrections to $\phi_p^{\text{LDA}}$ for
different semiconductors. The calculated Al/$X$~(100) SBH's including these
corrections are shown in the last two columns. All numbers are in eV.}
\begin{ruledtabular}
\begin{tabular}{cccc}
Semiconductor & Estimated &\multicolumn{2}{c}{$\phi_p$}\\
$X$ & correction & I & II\\
\hline
Ge & 0.17 & \multicolumn{2}{c}{0.21}\\
GaAs & 0.25 & 0.86 & 0.76\\
AlAs & 0.36 & 1.45 & 1.16\\
ZnSe & 0.36 & 2.18 & 1.82
\end{tabular}
\end{ruledtabular}
\end{table}

For the Al/AlAs system, Ref.~\onlinecite{Wang-83} reports SBH values for
Al/$n$-AlAs~(100) ranging from 0.85 to 0.94~eV for various reconstructions of
the semiconductor surface, while somewhat higher values, 0.95 and 1.01~eV, have
been given in Refs.~\onlinecite{Missous-90} and \onlinecite{Bosacchi-94},
respectively. Using the experimental bandgap $E_g^{\text{AlAs}}=2.16$~eV, the
resulting range is $\phi_p=1.15$--1.31~eV, in good agreement with the calculated
SBH for the Al-terminated AlAs surface (1.16~eV), and somewhat smaller than the
value we find for the As-terminated surface (1.45~eV). The Al/$n$-ZnSe~(100) SBH
has been investigated in Refs.~\onlinecite{Lazzarino-98}, \onlinecite{Chen-94},
and \onlinecite{Vos-89} for different reconstructions of the ZnSe~(100) surface.
Very similar values have been reported for the $c(2\times2)$ and $2\times1$
reconstructions, namely $\phi_p=2.12$--2.15~eV\cite{Lazzarino-98,Chen-94,Vos-89}
and $\phi_p=2.11$--2.15~eV,\cite{Lazzarino-98,Chen-94} respectively, while a
lower SBH, $\phi_p=1.91$~eV, has been measured for the $1\times1$
reconstruction.\cite{Lazzarino-98} These values are in between our values of
1.82~eV for the Zn-terminated and of 2.18~eV for the Se-terminated interface.

The general agreement between theory and experiment in Fig.~\ref{fig:phi_p-exp}
indicates that our calculations for ideal MS structures (abrupt interfaces with
no atomic relaxation) capture the general trend of the SBH with the chemical
composition of the semiconductor. We note that for Al/ZnSe, the inclusion of the
appropriate reconstruction and relaxation brings the theoretical results in very
close agreement with the experimental values.\cite{Lazzarino-98}

\section{Interpretation and models}

\subsection{General trend with the semiconductor chemical composition}

Experimentally a correlation was found between Schottky barriers and
heterojunction band offsets.\cite{Margaritondo-87} A large number of MS contacts
and semiconductor heterojunctions were shown to verify, within 0.4~eV, the
transitivity relationship:
	\begin{equation}\label{eq:transitivty}
		\phi_p[M/S_1]-\phi_p[M/S_2]=\Delta E_{\text{VBO}}[S_1/S_2],
	\end{equation}
where $M$ is a metal (such as Al or Au; in general neither a highly reactive
nor a transition metal\cite{Reactive}), $S_1$ and $S_2$ are two semiconductors,
and $\Delta E_{\text{VBO}}[S_1/S_2]$ is the corresponding valence-band offset
(VBO). This correlation was most often observed for MS junctions used in
transport measurements, i.e., which had been annealed for fabrication of the
contacts. The experimental data in Fig.~\ref{fig:phi_p-exp} are in general
agreement with the above empirical transitivity rule.

We note that the transitivity rule, as formulated in Eq.~(\ref{eq:transitivty}),
disregards any dependence of the SBH on the microscopic interface structure, and
cannot therefore give a complete account of the theoretical results in
Fig.~\ref{fig:phi_p-exp}. Also, recent theoretical and experimental studies have
shown that the band offset at heterovalent semiconductor heterojunctions depends
critically on the orientation and other microscopic details of the
interface.\cite{Baroni-89,Peressi-98,Franciosi-96} The right-hand side of
Eq.~(\ref{eq:transitivty}) is thus ill defined, in general, for heterovalent
semiconductors.

In this section we concentrate on the average SBH
$\bar{\phi}_p=\frac{1}{2}\left(\phi_p^{\text{I}} + \phi_p^{\text{II}}\right)$
of the abrupt, defect-free type-I and II interfaces, and propose a model for
its variation with the semiconductor chemical composition, derived from an
atomic-scale approach. We show that this variation is controlled essentially by
the same bulk mechanism that determines band offsets at non-polar, defect-free
semiconductor heterojunctions.\cite{Baroni-89} The splitting
$\Delta\phi_p=\phi_p^{\text{I}}-\phi_p^{\text{II}}$ due to the
semiconductor-surface termination will be the focus of the next section.

Similarly to the SBH, the VBO may be written as: $\Delta E_{\text{VBO}} =
\Delta E_v+\Delta V$, where $\Delta E_v$ is the difference between the VBM
energies of the two semiconductors, each measured relative to the mean
electrostatic potential in the corresponding crystal, and $\Delta V$ is the
electrostatic potential lineup at the interface. Since the band-structure terms
$\Delta E_v$ and $\Delta E_p$ [in Eq.~(\ref{eq:phi_p})] are differences between
bulk values of the individual crystals forming the junction, they verify by
definition the transitivity relationship in Eq.~(\ref{eq:transitivty}). All non
transitive contributions are contained thus in the potential lineup terms
$\Delta V$.

In the case of semiconductor heterojunctions a linear-response-theory (LRT)
approach, which focuses on $\Delta V$ and treats the interface as a
perturbation with respect to a bulk reference system, has provided an accurate
general description of band-offset trends.\cite{Baroni-89,Peressi-98} Based on
this approach and comparision with fully self-consistent {\it ab initio\/}
calculations, it has been shown, in particular, that in the case of
defect-free, isovalent lattice-matched semiconductor heterojunctions, $\Delta
V$ is determined by the properties of the bulk constituents (as opposed
to interface-specific features, such as interface orientation or interface
abruptness). Specifically, if $S_1$ and $S_2$ are the two semiconductors, with
anion (cation) species $a_1$ ($c_1$) in $S_1$ and $a_2$ ($c_2$) in $S_2$, the
potential lineup is given within LRT by:\cite{Baroni-89}
	\begin{equation}\label{eq:delta_V-model}
		\Delta V\left[S_1/S_2\right] = \frac{2\pi e^2}{3\Omega}
		\int r^2\left[\Delta n_a(\bm{r})+\Delta n_c(\bm{r})\right] d\bm{r},
	\end{equation}
where the integration is over the whole space, $\Omega$ is the volume of the
bulk unit cell, and $\Delta n_a$ ($\Delta n_c$) is the electronic charge
density induced by a single anion (cation) substitution $a_1 \rightarrow a_2$
($c_1 \rightarrow c_2$) in the bulk semiconductor $S_1$.\cite{origin} Based on
this LRT approach, it has also been shown that in the case of heterovalent
lattice matched semiconductors, Eq.~(\ref{eq:delta_V-model}) also applies in
the specific case of defect-free interfaces with the non-polar (110)
orientation.\cite{Peressi-98}

Using a similar linear-response scheme for $\Delta V$, we show in
Appendix~\ref{app:LRT} that the average SBH $\bar{\phi}_p$ can be described by
the following model:
	\begin{equation}\label{eq:phi_p-model}
		\bar{\phi}_p^{\text{mod}}=\phi_p\left[\text{Al}/\virtX~(100)\right] +
		\Delta E_{\text{VBO}}\left[\virtX/X~(110)\right].
	\end{equation}
The first term on the right-hand side of Eq.~(\ref{eq:phi_p-model}) is the SBH
at the (100) interface between Al and the group-IV virtual crystal, denoted
$\virtX$, which is obtained by averaging the anion and cation pseudopotentials
of the III-V or II-VI compound $X$ ($X=$ GaAs, AlAs, ZnSe). The second term is
the VBO of the non-polar $\virtX$/$X$~(110) heterojunction.

\begin{figure}[t!]
\includegraphics[width=8.6cm]{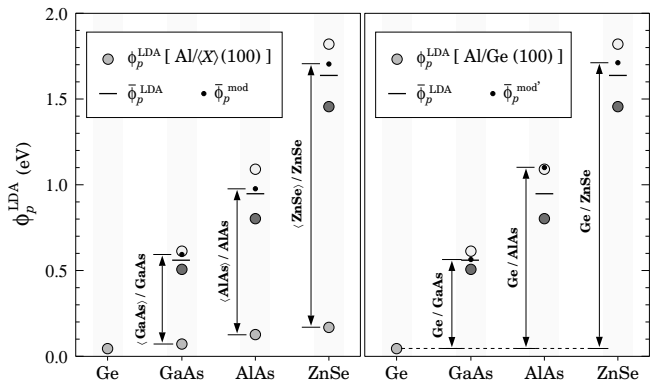}
\caption{\label{fig:phi_p-vbo} Comparison of the average SBH
$\bar{\phi}_p^{\text{LDA}}$ at the Al/$X$~(100) I and II interfaces with the
model predictions, Eq.~(\ref{eq:phi_p-model}) (left panel)
[Eq.~(\ref{eq:phi_p-model'}) (right panel)]. The horizontal bar shows the
average SBH and the small dot indicates the model result, i.e., the sum of the
SBH at the Al/$\virtX$~(100) [Al/Ge~(100)] junction (gray circles [dashed
line]) and the VBO at the $\virtX/X$~(110) [Ge/$X$~(110)] interface (double
arrows). The quasiparticle and spin-orbit corrections are not included; these
contributions trivially verify Eqs.~(\ref{eq:phi_p-model}) and
(\ref{eq:phi_p-model'}).}
\end{figure}

The basic approximation to derive Eq.~(\ref{eq:phi_p-model}) is to construct the
charge densities of the Al/$X$ I and II junctions (and hence their average
lineup) starting from the reference Al/$\virtX$ system, by adding a
linear superposition of the charge densities induced in the virtual crystal
$\virtX$ by single anion and cation substitutions that transform $\virtX$ into
$X$. The Al/$\virtX$~(100) junction is an optimal reference system in this
context, which minimizes the deviations of $\bar{\phi}_p$ from
$\bar{\phi}_p^{\text{mod}}$ in Eq.~(\ref{eq:phi_p-model}); these deviations
vanish to the first order in the ionic substitutions which transform Al/$\virtX$
into the Al/$X$ I and II junctions.

It is also possible to use as a reference system another Al/group-IV~(100)
junction, whose density is sufficiently close to the average density of the
Al/$X$ I and II junctions. For instance, one may use Al/Ge as a common
reference system and obtain (Appendix~\ref{app:LRT}):
	\begin{equation}\label{eq:phi_p-model'}
		\bar{\phi}_p^{\text{mod}'}=\phi_p\left[\text{Al}/\text{Ge}~(100)\right] +
		\Delta E_{\text{VBO}}\left[\text{Ge}/X~(110)\right].
	\end{equation}
The deviations of $\bar{\phi}_p$ from $\bar{\phi}_p^{\text{mod}'}$ in
Eq.~(\ref{eq:phi_p-model'}) include, in this case, a first order correction in
the substitutions. The latter correction can be identified with the dipole
induced in the reference Al/Ge~(100) junction by isovalent
Ge~$\rightarrow\virtX$ substitutions performed within the first one to three Ge
atomic layers closest to the interface (see Appendix~\ref{app:LRT}); such a
dipolar term is generally small for isovalent substitutions ($\sim 0.1$~eV or
less, see Ref.~\onlinecite{Bardi-99}), and will be neglected here.

In Fig.~\ref{fig:phi_p-vbo} we compare graphically the model predictions,
Eq.~(\ref{eq:phi_p-model}) and Eq.~(\ref{eq:phi_p-model'}), with the calculated
average SBH of the Al/$X$ I and II interfaces. The (110) VBO's have been
computed using supercells containing 8 planes of each semiconductor in the
ideal (unrelaxed) lattice-matched geometry. The same energy cutoffs and
$k$-points grids have been used as in the calculations of the Schottky
barriers. The SBH of the Al/$\virtX$~(100) junctions have been obtained using
the same parameters as for the Al/$X$~(100) I and II interfaces. The results in
Fig.~\ref{fig:phi_p-vbo} show that Eq.~(\ref{eq:phi_p-model}) and
Eq.~(\ref{eq:phi_p-model'}) provide a fairly accurate ($\pm 0.15$~eV)
description of the average SBH $\bar{\phi}_p$. We note that the SBH's at the
Al/$\virtX$ junctions are all small, due to the small bandgaps of the virtual
crystals ($<0.4$~eV), and similar to the LDA SBH at the Al/Ge~(100) interface
(0.04~eV). The results in Fig.~\ref{fig:phi_p-vbo} show that the (110) VBO ---
a bulk-related quantity in our calculations --- controls the general increase
of the barriers from the group-IV to the III-V and to the II-VI semiconductors.

\subsection{Effect of surface termination}
\label{sec:termination}

\begin{figure}[b!]
\includegraphics[width=8.6cm]{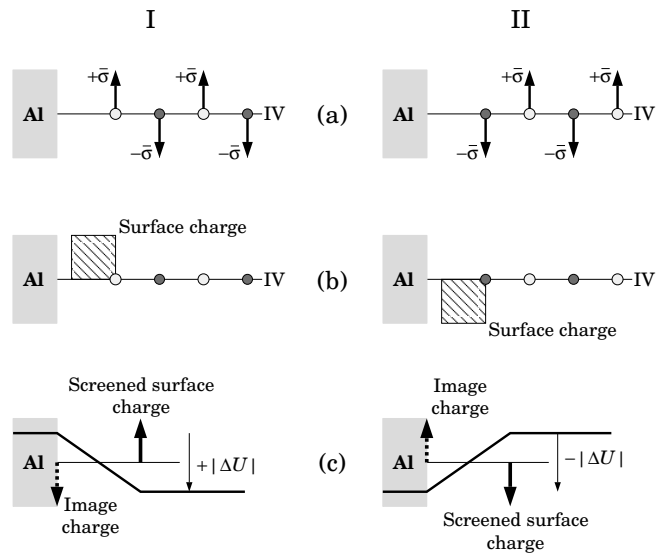}
\caption{\label{fig:image-charge}
(a) Planar average of the difference between the ionic charge densities of the
anion- (cation-)terminated Al/$X$~(100) interface and the Al/$\virtX$~(100)
interface; $\bar{\sigma}=1$ for the semiconductors $X=$ GaAs, AlAs, and
$\bar{\sigma}=2$ for $X=$ ZnSe. (b) Macroscopic average of the ionic charge
density difference. (c) Positive (negative) potential difference established at
the interface I (II) by a positive (negative) surface charge and its image
charge at the metal surface.}
\end{figure}

We will show here that the difference
$\Delta\phi_p=\phi_p^{\text{I}}-\phi_p^{\text{II}}$ due to the
semiconductor-surface termination in Fig.~\ref{fig:phi_p-exp} --- and in
particular the fact that the SBH is systematically higher for the anion than for
the cation termination --- can be understood in terms of surface-charge and
image-charge effects. The mechanism is illustrated in
Fig.~\ref{fig:image-charge}.

With respect to the Al/$\virtX$ interface, the ionic charge distributions of the
interfaces~I and II are obtained by substituting an anion (charge
$+\bar{\sigma}$) on each anionic site and a cation (charge $-\bar{\sigma}$) on
each cationic site, as indicated in Fig.~\ref{fig:image-charge}(a). For the
III-V and II-VI compounds we have $\bar{\sigma}=1$ and $\bar{\sigma}=2$,
respectively, in units of charge per unit-cell surface in the (100) plane. The
arrows in Fig.~\ref{fig:image-charge}(a) represent opposite delta functions on
each anionic and cationic (100) plane, corresponding to the planar average of
the ionic point-charge density. The macroscopic
average\cite{Baroni-89,Peressi-98} of this ionic charge distribution is
represented in Fig.~\ref{fig:image-charge}(b). In the bulk semiconductor, the
macroscopic average eliminates the atomic-scale oscillations of the planar
charge density; at the interface, however, a positive (negative) charge density
subsists in the junction I (II). This macroscopic charge has a density
$\rho=\pm2\bar{\sigma}/a$ and extends over a distance $a/4$ between the last Al
plane and the first semiconductor plane. It is therefore equivalent to a surface
charge of density $\sigma=\pm\bar{\sigma}/2$.

Within a classical macroscopic description, a plane of charge in a
semiconductor is screened by the dielectric constant $\epsilon_{\infty}$ of the
host material. Furthermore, in the presence of a metal the screened surface
charge is neutralized by an image charge induced at the metal surface, and a
potential difference is thus established between the two charges [see
Fig.~\ref{fig:image-charge}(c)]. If $\sigma$ is the density of surface charge
in the semiconductor, $x$ the position of the plane of surface charge, and
$x_i$ the position of the metal surface or image charge, the potential
difference obtained from classical electrostatics is:
	\begin{equation}\label{eq:deltaU-classical}
		\Delta U(x,\sigma) =4\pi e^2\frac{\sigma}{\epsilon_{\infty}}\,(x-x_i).
	\end{equation}
As can be seen from Fig.~\ref{fig:image-charge}(c), in the junction~I such a
dipole lowers the average potential energy in the semiconductor with respect to
its value in the metal, increasing the SBH $\phi_p$; conversely, in the
junction~II the dipole raises the average potential in the semiconductor,
decreasing $\phi_p$. The mechanism illustrated in Fig.~\ref{fig:image-charge}
thus provides a qualitative explanation for the difference between the SBH's of
the interfaces~I and II. Of course, the classical limit given by
Eq.~(\ref{eq:deltaU-classical}) is expected to be correct only for a charge
placed at a large distance from the metal. As we will see below, however,
closer to the metal the above type of description may still be used provided
the inhomogeneous nature of the screening near the metal is taken into account.

\begin{figure}[b!]
\includegraphics[width=8.6cm]{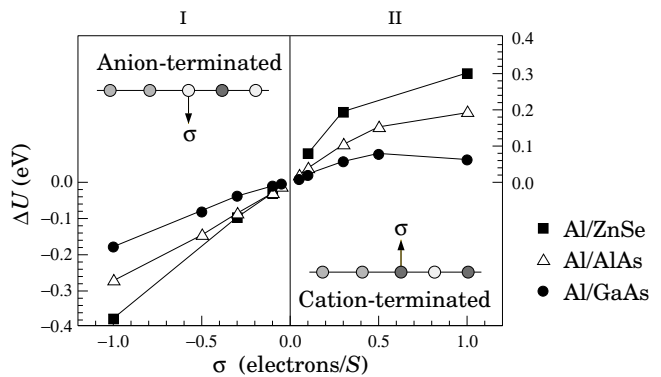}
\caption{\label{fig:deltaU-sigma}
Schottky barrier modification $\Delta U$ induced by a bare surface charge
$\sigma$ per unit-cell surface $S=a^2/2$ on the first semiconductor layer
closest to the metal in the Al/$X$~(100) I and II junctions (see insets; the
same symbols for the atomic layers are used as in Fig.~\ref{fig:structure}).}
\end{figure}

To check that the mechanism in Fig.~\ref{fig:image-charge} can indeed account
for the SBH differences $\Delta\phi_p$, we have calculated the changes in the
lineup (and hence in the SBH) induced by surface charges of varying magnitude,
placed on the semiconductor plane closest to the metal in the Al/$X$ I and II
junctions. At the interface I (II), a surface charge of density $-|\sigma|$
($+|\sigma|$) was introduced by replacing the anion A (cation C) of the
semiconductor layer adjacent to the metal surface by a virtual ion
$\langle\text{A}_{1-\frac{\sigma}{2}}\text{C}_{\frac{\sigma}{2}}\rangle$
($\langle\text{C}_{1-\frac{\sigma}{2}}\text{A}_{\frac{\sigma}{2}}\rangle$). The
resulting changes $\Delta U$ in the SBH obtained from the {\it ab initio\/}
calculations are shown in Fig.~\ref{fig:deltaU-sigma}. The negative (positive)
surface charge decreases (increases) the $p$-type SBH of the anion-terminated
(cation-terminated) interfaces, consistent with the screened surface charge and
image charge description in Fig.~\ref{fig:image-charge}(c). We also note that,
consistent with the latter description, the bare monopole is replaced by an
interface dipole in the multipole expansion of the total (electronic plus
ionic) charge disturbance.

The macroscopic average of the difference between the ionic potentials in the
junctions~II and I, in Fig.~\ref{fig:image-charge}(a), is equivalent to a
surface charge $\sigma=-1$ at the interface for the III-V semiconductors and
$\sigma=-2$ for the II-VI semiconductors. Therefore, focusing on the effect of
the macroscopic charges only and to the first order in the perturbation, the
modification of the SBH in the junctions~I for $\sigma=-1$ ($-2$) should be
equal to the difference $\phi_p^{\text{II}}-\phi_p^{\text{I}}$ for the III-V
(II-VI) semiconductors. Similarly, the change of the SBH in the junctions~II
induced by a surface charge $\sigma=+1$ ($+2$) should be equal to
$\phi_p^{\text{I}}-\phi_p^{\text{II}}$. Our {\it ab initio\/} results in
Fig.~\ref{fig:deltaU-sigma} show, however, that the responses $\Delta U$ of the
two interfaces are not linear when $|\sigma| \gtrsim 0.5 $ and differ in
magnitude. Therefore, we take the average $\bDU$ between the potential
differences induced in the junctions~I and II as our estimate for the
difference $\Delta\phi_p$. The results are shown in
Table~\ref{tab:delta_phi_p}. For Al/ZnSe we reported the calculated SBH changes
for $\sigma=\pm 2$ (not shown in Fig.~\ref{fig:deltaU-sigma}). The average
values $\bDU$ are seen to describe well the calculated difference
$\Delta\phi_p$ and also the increase of $\Delta\phi_p$ when the semiconductor
ionicity increases.

Although this supports the surface charge and image charge picture in
Fig.~\ref{fig:image-charge}(c), Eq.~(\ref{eq:deltaU-classical}) needs to be
revisited for charges placed close to the metal surface. For example, in the
case of a test charge $\sigma=0.1$ on the first semiconductor layer, we obtain
from Eq.~(\ref{eq:deltaU-classical}) a potential difference $\Delta U=82$~meV,
using for the distance $x-x_i$ the value $d/2$, where $d=1.72$~\AA\ is the
interplanar distance at the interface (see Fig.~\ref{fig:structure}), and for
$\epsilon_{\infty}$ the theoretical dielectric constant of GaAs,
$\epsilon_{\infty}^{\text{GaAs}}=12.4$.\cite{Berthod-96} This result is more
than 5 times larger than the {\it ab initio\/} result for the potential
difference obtained for Al/GaAs and shown in Fig.~\ref{fig:deltaU-sigma}.

In Ref.~\onlinecite{Berthod-96} we observed a somewhat similar behavior in the
case of local dipoles inserted in the Al/GaAs~(100) junction. In the latter
case, {\it ab initio\/} calculations were performed to determine the change in
the lineup $\Delta u$ induced by a dipole layer (i.e., two test charges
$+\sigma$ and $-\sigma$ placed on two adjacent cation-anion planes) introduced
at various distances $x$ from the metal, within the semiconductor. In this
numerical experiment, the bare dipole perturbation is $\Delta u_b = 4\pi
e^2\sigma l$, where $l$ is the separation between the charged planes, and from
the computed $\Delta u$ we could directly measure the effective dipole
screening $\epsilon_{\text{eff}}^{\text{dip}} = \Delta u_b/\Delta u$ as a
function of the dipole position $x$ in the junction. This screening was found
to be strongly inhomogeneous and to increase exponentially as the dipole was
approaching the metal surface. This was attributed to the MIGS tails and their
high polarizability in the interface region. We also proposed a model for
$\epsilon_{\text{eff}}^{\text{dip}}$ which proved very accurate to describe the
SBH changes $\Delta u(x,\,\sigma)$ in the linear-response regime (i.e., to the
first order in $\sigma$):\cite{Berthod-96}
	\begin{equation}\label{eq:screening-dipole}
		\epsilon_{\text{eff}}^{\text{dip}}(x)\approx\epsilon_{\infty}+
		4\pi e^2 D_s(E_{\text{F}},\,x)\,\delta_s.
	\end{equation}
Here $D_s(E_{\text{F}},\,x)$ is the MIGS surface density of states at the Fermi
energy and at the position $x$ of the dipole, and $\delta_s$ is the decay
length of the MIGS. The model Eq.~(\ref{eq:screening-dipole}) is also
consistent with earlier MIGS-based model descriptions of Schottky barrier
properties.\cite{Cowley-65,Louie-77,Zhang-85}

\begin{table}[t!]
\caption{\label{tab:delta_phi_p}
Comparison of the average SBH change $\bDUs$ induced by surface charges
$\sigma=\pm 1$ (GaAs, AlAs) and $\sigma=\pm 2$ (ZnSe) at the interface, with the
difference $\Delta\phi_p=\phi_p^{\text{I}}-\phi_p^{\text{II}}$ between the SBH
of the anion- and cation-terminated Al/$X$~(100) junctions (see text). The last
column shows the results of the model Eq.~(\ref{eq:delta_phi_p}). All energies
are in eV.}
\begin{ruledtabular}
\begin{tabular}{cccccccc}
$X$ & $\Delta\phi_p$ & \multicolumn{2}{c}{$|\Delta U|$} & $\bDUs$ & $D_s$ &
$\delta_s$ & $\Delta\phi_p^{\text{mod}}$\\
 & & I & II & & {\footnotesize (eV$^{-1}$\AA$^{-2}$)} & {\footnotesize (\AA)}
 &{\footnotesize Eq.~(\ref{eq:delta_phi_p})}\\
\hline
GaAs & 0.10 & 0.18 & 0.06 & 0.12 & 0.051 & 2.5 & 0.10\\
AlAs & 0.29 & 0.27 & 0.19 & 0.23 & 0.060 & 2.0 & 0.13\\
ZnSe & 0.36 & 0.68 & 0.24 & 0.46 & 0.041 & 1.8 & 0.40\\
\end{tabular}
\end{ruledtabular}
\end{table}

In order to predict, in general, the effect of the surface termination in MS
junctions, we would like to develop a model for $\Delta U$ that takes into
account the inhomogeneous nature of the electronic screening in the MIGS region
and that is consistent with our previous results on the effect of the dipole
layers on the SBH. In particular, this model should be consistent with the fact
that, in the linear-response regime (small $|\sigma|$), the sum of the SBH
modifications induced separately by two charges $+\sigma$ and $-\sigma$
separated by a small distance $l$, $\Delta U(x-l/2,\,\sigma)$ and $-\Delta
U(x+l/2,\,\sigma)$, respectively, must be equal to the SBH modification induced
by the corresponding dipole:
	\begin{equation}\label{eq:charges-dipole}
		\Delta U(x-l/2,\,\sigma)-\Delta U(x+l/2,\,\sigma)=
		-\frac{4\pi e^2\sigma l}{\epsilon_{\text{eff}}^{\text{dip}}(x)}.
	\end{equation}
Expanding the left-hand side of Eq.~(\ref{eq:charges-dipole}) to the first
order in $l$, we obtain the differential equation:
	\begin{equation}\label{eq:U-derivative}
		\frac{\partial\Delta U}{\partial x} (x,\,\sigma) =
		\frac{4\pi e^2\sigma}{\epsilon_{\text{eff}}^{\text{dip}}(x)}.
	\end{equation}
With our expression for $\epsilon_{\text{eff}}^{\text{dip}}(x)$ in
Eq.~(\ref{eq:screening-dipole}) and a surface density of states that decays
exponentially,\cite{exponential}
$D_s(E_{\text{F}},\,x)= D_s(E_{\text{F}},\,0)\exp(-x/\delta_s)$,
the solution of Eq.~(\ref{eq:U-derivative}) with the boundary condition $\Delta
U(x_0)=0$ is:
	\begin{equation}\label{eq:deltaU-MIGS}
		\Delta U(x,\,\sigma)=4\pi e^2\frac{\sigma}{\epsilon_{\infty}}
		\left[x-x_0-\delta_s\log\frac{\epsilon_{\text{eff}}^{\text{dip}}(x_0)}
		{\epsilon_{\text{eff}}^{\text{dip}}(x)}\right],
	\end{equation}
where $x$ is the position of the surface charge. We note that for large values
of $x$, we recover the classical limit given by Eq~(\ref{eq:deltaU-classical}),
with $x_i = x_0 + \delta_s
\log[\epsilon_{\text{eff}}^{\text{dip}}(x_0)/\epsilon_{\infty}]$.

With the exception of $x_0$, all parameters necessary to evaluate $\Delta
U(x,\,\sigma)$ in Eq.~(\ref{eq:deltaU-MIGS}) can be obtained straightforwardly
from {\it ab initio\/} calculations performed either for the bulk semiconductor
($\epsilon_{\infty}$) or for the unperturbed Al/$X$ junction (the MIGS-related
parameters). In Table~\ref{tab:delta_phi_p}, we have reported our calculated
values for the MIGS parameters $D_s \equiv D_s(E_{\text{F}},\,0)$ and
$\delta_s$. These quantities were obtained from the calculated macroscopic
average of the local density of states, $N(E,\,x)$, as
$D_s=\int_0^{\infty}N(E_{\text{F}},\,x)\,dx$ and
$\delta_s=\frac{1}{D_s}\int_0^{\infty}x\,N(E_{\text{F}},\,x)\,dx$, where the
origin ($x=0$) was taken as the midpoint between the last Al and the first
semiconductor plane, and $\infty$ indicates a position well inside the
semiconductor (the center of the semiconductor slab in the supercell) where the
MIGS vanish. As the values of $D_s$ and $\delta_s$ are slightly different for
the interfaces~I and II, we reported in Table~\ref{tab:delta_phi_p} the average
between the values calculated for the two interfaces.\cite{Previous_delta_s}

\begin{figure}[t!]
\includegraphics[width=8.6cm]{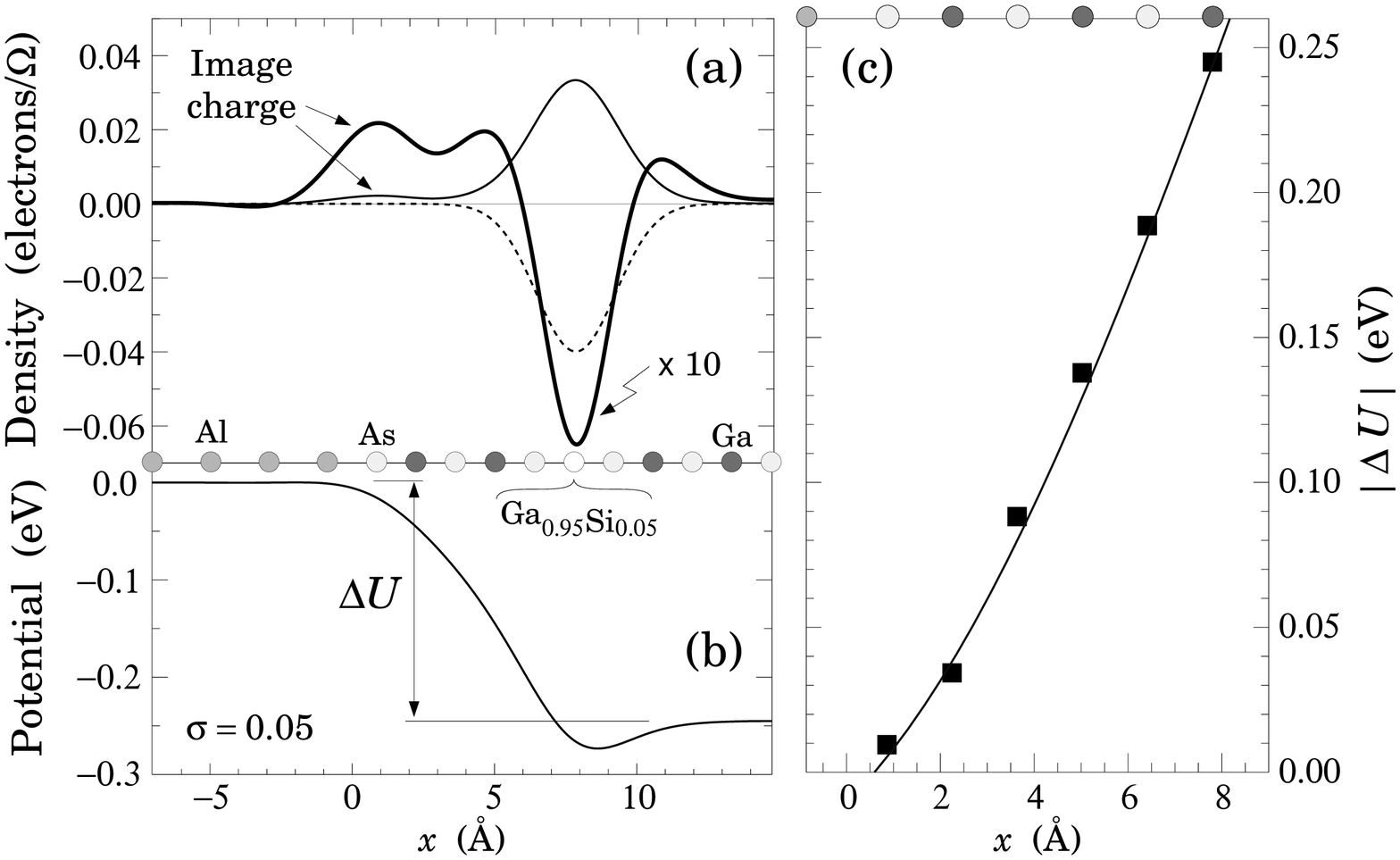}
\caption{\label{fig:dipole-charge}
(a) Macroscopic average of the electronic (thin solid line) and ionic (dotted
line) charge densities induced by a plane of
$\langle\text{Ga}_{0.95}\text{Si}_{0.05}\rangle$ virtual ions in the
As-terminated Al/GaAs~(100) junction. A Gaussian filter function was used for
the macroscopic average. The thick solid line is the sum of the electronic and
ionic densities, scaled by a factor of 10. (b) Macroscopic average of the
corresponding induced total electrostatic potential. The resulting potential
difference $\Delta U$ is also indicated. (c) Schottky barrier modification
$|\Delta U|$ (filled squares) obtained for a surface charge $|\sigma|=0.05$ as
a function of its position within the semiconductor, in the As-terminated
Al/GaAs junction. The symbols give the results of the self-consistent
calculations. The solid line corresponds to the prediction of
Eq.~(\ref{eq:deltaU-MIGS}) with $x_0=0.6$~\AA. The atomic positions are
indicated using the same symbols for the atoms as in Fig.~\ref{fig:structure}.
The calculations were done in a $7+21$ supercell.}
\end{figure}

In order to obtain an estimate for $x_0$, and also to test the model in
Eq.~(\ref{eq:deltaU-MIGS}), we have investigated {\it ab initio\/} the spatial
dependence of $\Delta U$ in the linear-response regime by introducing a small
test surface charge $\sigma=\pm0.05$ in the As-terminated Al/GaAs junction at
different distances from the interface. This was done by replacing single
layers of As (Ga) ions by virtual
$\langle\text{As}_{0.95}\text{Si}_{0.05}\rangle$
($\langle\text{Ga}_{0.95}\text{Si}_{0.05}\rangle$) anions (cations). As an
example, we show in Fig.~\ref{fig:dipole-charge} the {\it ab initio\/} results
for the charge density and potential induced by such a test charge on the sixth
semiconductor layer from the metal. The macroscopic averages of the ionic,
electronic, and total charge densities are displayed in
Fig.~\ref{fig:dipole-charge}(a). We have used a Gaussian filter function with
full width at half maximum $a/2$ for the macroscopic average. This allows one,
in particular, to distinguish the image charge contribution to the total charge
density, close to the Al surface. The macroscopic average of the induced total
electrostatic potential is displayed in Fig.~\ref{fig:dipole-charge}(b). The
corresponding potential difference is $\Delta U=0.24$~eV.

In Fig.~\ref{fig:dipole-charge}(c) we have plotted the discontinuity $|\Delta
U|$ induced by the test charge as a function of its position in the Al/GaAs
junction. With the theoretical dielectric constant of GaAs
($\epsilon_{\infty}^{\text{GaAs}}=12.4$) and the calculated values of $D_s$ and
$\delta_s$ given in Table~\ref{tab:delta_phi_p}, the best fit of
Eq.~(\ref{eq:deltaU-MIGS}) to the data in Fig.~\ref{fig:dipole-charge}(c) is
obtained with $x_0=0.6$~\AA. The model results obtained from
Eq.~(\ref{eq:deltaU-MIGS}) using $x_0=0.6$~\AA\, have been reported in
Fig.~\ref{fig:dipole-charge}(c), and compare well with the results of the
self-consistent calculations as a function of the distance.

Having a reasonable estimate for $x_0$ we may now use
Eq.~(\ref{eq:deltaU-MIGS}) to obtain also an estimate for the difference
between the SBH's of the interfaces~I and II. As we have seen before, this
difference may be evaluated as the potential change induced, to the first
order, by a surface charge $\bar{\sigma}=1$ ($2$) per unit-cell surface on the
first plane of the III-V (II-VI) semiconductor, i.e., at a position $x=d/2$
with $d=1.72$~\AA\ (see Fig.~\ref{fig:structure}). The resulting estimate
$\Delta\phi_p^{\text{mod}}$ for the difference $\Delta\phi_p$ is thus
	\begin{equation}\label{eq:delta_phi_p}
		\Delta\phi_p^{\text{mod}} = 4\pi e^2\frac{\bar{\sigma}}
		{\epsilon_{\infty}}\left[d/2-x_0-
		\delta_s\log\frac{\epsilon_{\text{eff}}^{\text{dip}}(x_0)}
		{\epsilon_{\text{eff}}^{\text{dip}}(d/2)}\right].
	\end{equation}
The results obtained with this model are displayed in
Table~\ref{tab:delta_phi_p}. To evaluate $\Delta\phi_p^{\text{mod}}$, we have
used the theoretical value of the semiconductor dielectric constant
$\epsilon_{\infty}^X$ ($X=$ GaAs, AlAs, and ZnSe),\cite{diel} the surface
density of states $D_s$ and MIGS decay length $\delta_s$ given in
Table~\ref{tab:delta_phi_p} for the Al/GaAs, Al/AlAs and Al/ZnSe junctions, and
$x_0=0.6$~\AA\ for all systems. The model results, in
Table~\ref{tab:delta_phi_p}, yield the correct trend and order of magnitude for
the difference between the SBH's. We note that decreasing $x_0$ increases
$\Delta\phi_p^{\text{mod}}$, but does not affect the trend. The same conclusion
applies when a possible change in $x_0$, from GaAs to AlAs and to ZnSe, is
taken into account by scaling the $x_0$ value obtained for GaAs with the ratio
of the MIGS decay length, i.e.,
$\delta_s^X/\delta_s^{\text{GaAs}}$.\cite{x0_depend}

Although the present model approach provides a consistent picture of the effect
of selected perturbations at MS interfaces, we would like to caution the reader
that our model description applies to unrelaxed interfaces. In this connection
it should be pointed out that a LRT model (based on dynamical effective
charges) is available in the literature for quantitative predictions of the
atomic relaxation contribution to the SBH.\cite{Ruini-97}

\section{Conclusion}

Using a first-principle pseudopotential approach, we have investigated the
Schottky barrier heights of abrupt lattice-matched Al/Ge, Al/GaAs, Al/AlAs, and
Al/ZnSe~(100) junctions, and their dependence on the semiconductor chemical
composition and surface termination. The sensitivity of the SBH to microscopic
interface features reveals the limits of the currently accepted semiempirical
models of Schottky barriers. Such model theories generally neglect the effects
of the microscopic interfacial morphology. This is due in part to the
complexity of the actual atomic structure of most MS contacts, and also to the
relatively limited information available on the atomic-scale geometry of buried
interfaces.

Based on our {\it ab initio\/} studies, we have derived models which explicitly
include the effects of the interface atomic structure in the case of some
prototype defect-free, lattice-matched geometries. These models retain, within
specific ranges of applicability, the same accuracy as the {\it ab initio\/}
calculations. They show, in particular, that while the variation of the average
SBH of the abrupt, anion- and cation-terminated Al/semiconductor~(100)
interfaces can be explained mainly in terms of the bulk properties of the
semiconductors, the difference between the barrier heights of the anion- and
cation-terminated interfaces results from a microscopic dipole generated by the
screened charge of the polar semiconductor surface and its image charge at the
metal surface. Our atomic-scale computations have also allowed us to show how
the classical image charge concept, valid for charges placed at large distances
from the metal, can be extended to distances shorter than the decay length of
the metal-induced-gap states.

\begin{acknowledgments}

One of us (CB) acknowledges support by the Swiss National Science Foundation
under Grant N$^{\circ}$ 20-47065.96. The computations were performed at the
CSCS in Manno.

\end{acknowledgments}

\begin{table*}[t!]
\caption{\label{tab:transitivity}
Comparison of the average potential lineup $\bDV$ at the Al/$X$~(100) I and II
interfaces with the predictions of the model, $\bDV \approx \Delta
V\left[\text{Al}/\virtY~(100)\right] + \Delta V\left[\virtY/X~(110)\right]$
[Eq.~(\ref{eq:deltaV-LRT})], using $\virtY=\virtX$ ($\bDV^{\text{mod}}$) and
$\virtY=\text{Ge}$ ($\bDV^{\text{mod}'}$). All numbers are in eV.}
\begin{ruledtabular}
\begin{tabular}{cccccccc}
$X$ & $\bDV$ & \multicolumn{2}{c}{$\Delta V$} & $\bDV^{\text{mod}}$ &
\multicolumn{2}{c}{$\Delta V$} & $\bDV^{\text{mod}'}$\\
 & & Al/$\virtX$~(100) & $\virtX/X$~(110) & & Al/Ge~(100) & Ge/$X$~(110) &\\
\hline
GaAs & $-2.34$ & $-2.18$ & $-0.12$ & $-2.30$ & $-2.05$ & $-0.28$ & $-2.33$\\
AlAs & $-2.01$ & $-1.97$ & $-0.01$ & $-1.98$ & $-2.05$ & $+0.20$ & $-1.85$\\
ZnSe & $-3.30$ & $-2.85$ & $-0.38$ & $-3.23$ & $-2.05$ & $-1.17$ & $-3.22$\\
\end{tabular}
\end{ruledtabular}
\end{table*}

\appendix

\section{Model for the average SBH of the anion- and cation-terminated
interfaces}\label{app:LRT}

In order to explain the behavior of the average SBH $\bar{\phi}_p$ with the
semiconductor composition, we have extended to MS contacts a
linear-response-theory approach commonly used to study band offsets at
semiconductor heterojunctions.\cite{Baroni-89,Peressi-98} The present analysis
is also an extension to {\it heterovalent\/} semiconductors of an approach
outlined in Ref.~\onlinecite{Bardi-99} to model the Schottky-barrier changes
with the alloy composition in Al/Ga$_{1-x}$Al$_x$As junctions.

We are interested in the average potential lineup $\bDV=\frac{1}{2}\left(\Delta
V^{\text{I}}+\Delta V^{\text{II}}\right)$, where $\Delta V^{\text{I(II)}}$ is
the potential lineup at the interface~I (II), for which we want to establish the
following model:
	\begin{equation}\label{eq:deltaV-LRT}
		\bDV \approx \Delta V\left[\text{Al}/\virtY~(100)\right]+
		\Delta V\left[\virtY/X~(110)\right].
	\end{equation}
The first term on the right-hand side of Eq.~(\ref{eq:deltaV-LRT}) is the
potential lineup at the reference (100) junction between Al and a group-IV (real
or virtual) crystal $\virtY$ (e.g., the virtual crystal $\virtX$ or Ge), having
a charge density close to the average charge density of the Al/$X$ I and II
junctions. The second term is the lineup at the non-polar (110) interface
between the group-IV crystal $\virtY$ and the semiconductor $X$. To derive
Eq.~(\ref{eq:deltaV-LRT}), we write the self-consistent electrostatic potential
at the Al/$X$~(100) I (II) junction as
	\begin{equation}\label{eq:potential-I-II}
		V^{\text{I(II)}}(\bm{r}) = V_0(\bm{r}) + V_1^{\text{I(II)}}(\bm{r}),
	\end{equation}
where $V_0(\bm{r})$ is the electrostatic potential at the Al/$\virtY$ (100)
junction. The average lineup $\bDV$ can be expressed, according to
Eq.~(\ref{eq:potential-I-II}), as $\bDV=\Delta V_0+\Delta V_1$, where $\Delta
V_0\equiv\Delta V\left[\text{Al}/\virtY~(100)\right]$ and $\Delta V_1$ is the
lineup of the potential
	\begin{equation}\label{eq:potential1}
		V_1(\bm{r}) =\frac{1}{2}\left[V_1^{\text{I}}(\bm{r})+
		V_1^{\text{II}}(\bm{r})\right].
	\end{equation}
The potential $V_1^{\text{I(II)}}$ is the self-consistent electrostatic
potential induced in the Al/$\virtY$~(100) junction by the ionic substitutions,
$\virtY \rightarrow$ anion and $\virtY \rightarrow$ cation, performed in the
semiconductor, which transform the Al/$\virtY$ system into the type-I (type-II)
Al/$X$ system. Thus $V_1^{\text{I}}$ and $V_1^{\text{II}}$ have long-range
contributions associated with each heterovalent anion and cation substitution in
the group-IV crystal. These long-range terms cancel out in the average in
Eq.~(\ref{eq:potential1}), since each anion (cation) substitution in
$V_1^{\text{I}}$ is compensated by a cation (anion) substitution associated with
the {\em same site\/} in $V_1^{\text{II}}$. The average potential $V_1$ has
therefore a well defined macroscopic average in the semiconductor, which is
equal to $\Delta V_1$, since $V_1(\bm{r})$ vanishes in the metal.

One may thus evaluate $\Delta V_1$ using a perturbative approach neglecting
inter-site interactions in the ionic substitutions, because of the short-ranged
nature of the potentials associated with each individual site. Within this
approximation, $V_1$ is given by {\it i\/}) the superposition of the potentials
induced by {\em isolated\/} anion and cation substitutions in the bulk crystal
$\virtY$ plus {\it ii\/}) a correction due to the deviations from the bulk
response for substitutions performed near the MS interface.

By construction, the potential lineup obtained from {\it i\/}) is orientation
independent and equal to the potential lineup $\Delta
V'\left[\virtY/X~(110)\right]$ at the non-polar $\virtY/X$~(110) interface,
built from a superposition of the same isolated charge-density responses on one
side of the $X$~(110) homojunction. Furthermore, previous {\it ab initio\/} and
LRT studies of semiconductor heterojunctions have shown that the deviation of
the lineup $\Delta V$~(110) from the transitivity rule, and hence from $\Delta
V'$~(110), is typically less than 0.1~eV in IV-IV/III-V junctions, and of the
order of 0.1~eV in IV-IV/II-VI junctions.\cite{Baroni-89,Peressi-98} We may
therefore replace $\Delta V'\left[\virtY/X~(110)\right]$ by $\Delta
V\left[\virtY/X~(110)\right]$ to obtain the contribution from {\it i\/}) to
$\Delta V_1$.

The correction to the lineup induced by {\it ii\/}) is given, to the first
order in the substitutions, by $\Delta V_{\text{corr.}} = \sum_i d_i$, where
$d_i = 4 \pi e^2 \int dx\,x\Delta n_i(x)$ is the dipole, and $n_i(x)$ the
charge density, induced by the $\virtY \rightarrow$ anion or $\virtY
\rightarrow$ cation layer substitution within the $i^{\text{th}}$ atomic plane
from the interface in the Al/$\virtY$ junction. In practice, the dipoles $d_i$
vanish beyond the $3^{\text{rd}}$ to $4^{\text{th}}$ atomic plane from the
junction, and $\Delta V_{\text{corr.}}$ is generally of the order of 0.1~eV for
isovalent substitutions.\cite{Bardi-99} Furthermore, when the Al/$\virtX$
junction is used as a reference system, $\Delta V_{\text{corr.}}$ exactly
vanishes, because the corrections are opposite in the I and II junctions and
cancel out in the average leading to $\Delta V_1$. The correction $\Delta
V_{\text{corr.}}$ is therefore bound to be small ($\sim0.1$~eV) when the
reference system is an Al/$\virtY$~(100) junction with a density close to the
average density of the Al/$X$ I and II junctions. We will thus neglect this
correction, which leads to $\Delta V_1 \approx \Delta
V\left[\virtY/X~(110)\right]$, and hence to Eq.~(\ref{eq:deltaV-LRT}).

In Table~\ref{tab:transitivity}, the average potential lineup at the
Al/$X$~(100) I and II interfaces is compared to the predictions of the model,
Eq.~(\ref{eq:deltaV-LRT}), obtained with $\virtY=\virtX$ and with
$\virtY=\text{Ge}$. The agreement between $\bDV^{\text{mod}}$
[$\bDV^{\text{mod}'}$] and the calculated $\bDV$ is quite good, the discrepancy
being 2\% [8\%] or less when the Al/$\virtX$~(100) [Al/Ge~(100)] junction is
used as a reference system. Introducing the band energies in
Eq.~(\ref{eq:deltaV-LRT}), we obtain Eq.~(\ref{eq:phi_p-model})
[Eq.~(\ref{eq:phi_p-model'})] with $\virtY=\virtX$ [$\virtY=\text{Ge}$].


\begin{thebibliography}{99}
\bibitem{Pseudo} The Ga~${3d}$ and Zn~${3d}$ electrons were treated as frozen
core orbitals. It has been shown that including the Ga~${3d}$ orbitals in the
valence shell amounts to including a rigid (semiconductor bulklike) correction
of $-0.1$~eV on the Al/GaAs SBH, which we neglect here.
\bibitem{erratum} In Ref.~\onlinecite{Berthod-98} the many-body and spin-orbit
correction used for the Al/GaAs SBH was 0.21~eV, and not 0.11~eV as indicated
by mistake in the caption of Table~I.
\bibitem{origin} It should be noted that $\Delta n_a(\bm{r})$ [$\Delta
n_c(\bm{r})$] in Eq.~(\ref{eq:delta_V-model}) is neutral, and carries no dipole
nor quadrupole (except for the second spherical moment of the charge given by
Eq.~(\ref{eq:delta_V-model})), because of the tetrahedral $T_d$ symmetry of the
impurity site (see Ref.~\onlinecite{Baroni-89} or \onlinecite{Peressi-98}). The
integral in Eq.~(\ref{eq:delta_V-model}) is therefore well defined; in
particular, the result does not depend on the choice of the origin.
\bibitem{Previous_delta_s} We note that a somewhat larger value of $\delta_s$
(3.2~\AA) was used in Ref.~\onlinecite{Berthod-96} to model the effect of Si
interlayers on the Al/GaAs SBH. In the latter study, $\delta_s$ was not
calculated from first principles, but taken as a parameter value giving good
general agreement with the SBH modifications induced in a junction containing
up to 2 Si monolayers. The estimated uncertainty on the present $\delta_s$
calculated for the unperturbed Al/$X$ junction is about $\pm 0.2$~\AA. This
uncertainty takes into account deviations of $N(E_{\text{F}},\,x)$ from an
ideal exponential function, which give rise to some variations of $\delta_s$
with $x$. Such a change in $\delta_s$, however, has a small influence on the
model results in Fig.~\ref{fig:dipole-charge}(c) and
Table~\ref{tab:delta_phi_p} (12\% at most).
\end{thebibliography}
\end{document}